\documentclass[12pt]{iopart}

\usepackage[svgnames]{xcolor}
\usepackage{graphicx}
\usepackage{dcolumn}
\usepackage{bm}

\usepackage{amssymb}

\newcommand{\virg}[1]{``#1''}

\begin{document}

\title{Low-temperature spectroscopy of the $^{12}$C$_2$H$_2$ ($\upsilon_1 +\upsilon_3$) band in a helium buffer gas}

\author{L.~Santamaria$^{1}$, V.~Di~Sarno$^{1}$, I.~Ricciardi$^{1}$, M.~De~Rosa$^{1}$, S.~Mosca$^{1}$, G.~Santambrogio$^{2,3}$, P.~Maddaloni$^{1,4}$, and P.~De~Natale$^{4,5}$}

\address{$^{1}$ CNR-INO, Istituto Nazionale di Ottica, Via Campi Flegrei 34, 80078 Pozzuoli, Italy}
\address{$^{2}$ CNR-INO, Istituto Nazionale di Ottica, Via N. Carrara 1, 50019 Sesto Fiorentino, Italy}
\address{$^{3}$ Fritz-Haber-Institut der Max-Planck-Gesellschaft, Faradayweg 4-6, 14195 Berlin, Germany}
\address{$^{4}$ INFN, Istituto Nazionale di Fisica Nucleare, Sez. di Firenze, Via G. Sansone 1, 50019 Sesto Fiorentino, Italy}
\address{$^{5}$ CNR-INO, Istituto Nazionale di Ottica, Largo E. Fermi 6, 50125 Firenze, Italy}

\ead{pasquale.maddaloni@ino.it}
\vspace{10pt}
\begin{indented}
\item[]October 2014
\end{indented}

\begin{abstract}
Buffer gas cooling with a $^4$He gas is used to perform laser-absorption spectroscopy of the $^{12}$C$_2$H$_2$ ($\nu_1+\nu_3$) band at cryogenic temperatures. Doppler thermometry is first carried out to extract translational temperatures from the recorded spectra. Then, rotational temperatures down to 20 K are retrieved by fitting the Boltzmann distribution to the relative intensities of several ro-vibrational lines. The underlying helium-acetylene collisional physics, relevant for modeling planetary atmospheres, is also addressed. In particular, the diffusion time of $^{12}$C$_2$H$_2$ in the buffer cell is measured against the $^4$He flux at two separate translational temperatures; the observed behavior is then compared with that predicted by a Monte Carlo simulation, thus providing an estimate for the respective total elastic cross sections: $\sigma_{el}(100\ {\rm K})=(4\pm1)\cdot 10^{-20}$ m$^{2}$ and $\sigma_{el}(25\ {\rm K})=(7\pm2)\cdot 10^{-20}$ m$^{2}$. 
\end{abstract}


\vspace{2pc}

\submitto{The Astrophysical Journal}
%
%
%

\section{\label{sec:level1}Introduction}
By virtue of its prototypical role in different research areas, acetylene has been the subject of extensive spectroscopic studies \cite{Herman2003, Herman2007}. First, the paradigmatic carbon-carbon triple bond provides a fertile ground for the exploration of fundamental quantum chemistry processes in molecular beams, including reactions and collisions as well as the formation of van der Waals complexes \cite{Thibault2007, Thorpe2009, Didriche2012}. From a technological perspective, much work in the field of high-resolution spectroscopy has been motivated by the demand for improved frequency standards and metrological capabilities in the telecom spectral region \cite{Edwards2005, Hardwick2006, Ryu2008, Ahtee2009}. Moreover, trace-molecule spectroscopy of acetylene is of considerable interest in atmospheric chemistry and geophysical research in connection with pollution control and global climate, respectively \cite{Rinsland1987}. While representing only a trace component on Earth, acetylene is formed, by photolysis of methane, in the atmospheres of jovian planets (Jupiter, Saturn, Uranus, and Neptune) and Titan, as well as in various other stellar and interstellar environments \cite{Varanasi1983, Noll1986, Conrath1989}; as such, acetylene is also a key species in astrophysics and astrobiology \cite{Oremland2008}. 

Potentially profitable in all the above applications, laboratory spectroscopy investigations of acetylene in the low-temperature regime are crucial to understand and model planetary atmospheres. Indeed, it was thanks to the $^{12}$C$_2$H$_2$ ro-vibrational emission spectra at 13.7 $\mu$m ($\upsilon_5$-fundamental band) observed by the instruments on board \textit{Voyagers 1} and \textit{2} that the atmospheric temperature of Jupiter (about 130 K), Titan (between 120 and 130 K), Saturn (around 90 K) and Neptune (below 60 K) were retrieved \cite{Varanasi1992}. More recently, infrared spectroscopic measurements performed by the Spitzer Space Telescope discovered trace amounts of acetylene in the troposphere of Uranus as well, consistent with a lowest recorded temperature of 49 K \cite{Burgdorf2006}. While spectral lines in planetary atmospheres are mainly influenced by collisions with molecular hydrogen, atomic helium plays an important role, too. In this regard, calculations and measurements of collisional broadening and shift coefficients were specifically carried out for the helium-acetylene system, first on the mid-infrared $(\upsilon_4 +\upsilon_5 )$ \cite{Podolske1984} and $\upsilon_5$ \cite{Bouanich1991, Varanasi1992, Babay1998, Heijmen1999} bands, respectively at 7.4 and 13.7 $\mu$m, and then on the near-infrared $(\upsilon_1 +3\upsilon_3)$ \cite{Valipour2001} and $(\upsilon_1 +\upsilon_3)$ \cite{Thibault2005, Arteaga2007, Bond2008} bands, at 788 nm and 1.5 $\mu$m, respectively; however, most of these studies focused on room-temperature systems, except for a couple of works reporting temperatures just below 195 \cite{Bond2008} and 150 \cite{Podolske1984} K. The general difficulty encountered in accessing the range of tens of Kelvin with laboratory spectroscopic setups lies in the fact that most of the species of interest, including acetylene, have poor vapor pressure in that temperature interval. Only in very few cases, based on a special collisional cooling methodology, was such a limitation overcome and significantly lower temperatures, down to 4 K, reached \cite{Messer1984}. This allowed comprehensive investigations of pressure broadening in the CO-He system \cite{Messer1984}, He-induced rotational relaxation of H$_2$CO \cite{Ball1998}, and rotational inelastic cross sections for H$_2$S-He collisions \cite{Mengel2000}. Nevertheless, this approach has never been applied to acetylene. 

A new impetus to this research line comes from the emerging, powerful technologies for the cooling of stable molecules \cite{Carr2009}. Among the various schemes, at least for temperatures in the few-Kelvin range, the buffer-gas-cooling (BGC) method is perhaps the most efficient in terms of produced sample density and it is applicable to nearly all molecules \cite{Maxwell2005, Bulleid2013}. Here, a noble gas, typically helium, is chilled just above its boiling point and acts as a thermal bath (buffer) that cools in turn, through collisions, the injected molecular gas under analysis. 

In this work, a BGC apparatus with $^4$He (boiling point of $\simeq$ 4.2 K at 1 atm) is used to prepare a $^{12}$C$_2$H$_2$ (boiling point of $\simeq$ 190 K at 1 atm) sample at temperatures which are characteristic of planetary atmospheres and the interstellar medium. In this regime, laser absorption spectroscopy is performed, primarily aimed at determining, in conjunction with the outcomes of a Monte Carlo simulation, the total (as opposed to differential) elastic cross sections for the $^4$He$-^{12}$C$_2$H$_2$ system. For this purpose, a thorough characterization of the BGC process is first accomplished, comprising measurements of translational temperatures by means of Doppler thermometry, as well as of internal (rotational) temperatures through the analysis of the relative intensities of several rotational lines. 
 
\section{Experimental setup}
\begin{figure}[h!]
\centering
\includegraphics[width=\textwidth]{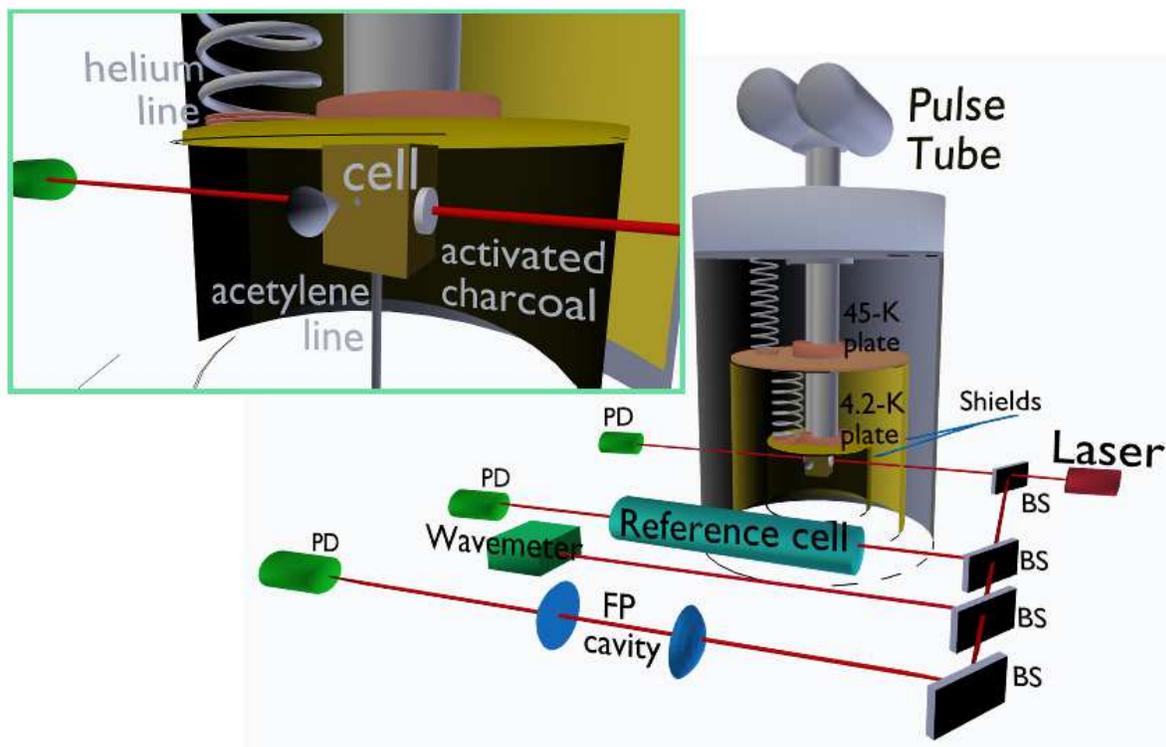}
\caption{Layout of the experimental setup including a zoom on the buffer-gas cell. Laser absorption spectroscopy is used to characterize collisional cooling of $^{12}$C$_2$H$_2$ in a $^4$He thermal bath down to a temperature of few Kelvin.}\label{setup}
\end{figure}

Described in detail in a previous work \cite{Santamaria2014}, the heart of the experimental apparatus is represented by a two-stage pulse tube (PT) cryocooler (Cryomech, PT415) housed in a stainless-steel vacuum chamber and fed with liquid helium by a compressor. The first (second) PT stage yields a temperature of 45 K (4.2 K) provided that its heat load is kept below 40 W (1.5 W); to guarantee this, each plate is enclosed in a round gold-plated copper shield, which suppresses black-body radiation effects. Capillary filling, regulated upstream by two flow controllers with an accuracy of 0.05 Standard Cubic Centimeters per Minute (SCCM), is used to inject both acetylene and helium, contained in room-temperature bottles, into the buffer cell. This latter consists of a gold-plated copper cube of side length $L_{c}=54$ mm; it is in thermal contact with the 4.2-K plate and its exit hole has a radius of $r_h=1$ mm. The acetylene pipe is made of stainless-steel and thermally insulated from both the PT stages; in addition, to avoid condensation, its temperature is maintained above 190 K by means of a proportional-integral-derivative (PID) feedback loop equipped with a silicon-diode thermometer as input sensor and an electric heater as output transducer. The buffer gas line comprises four connected segments: the first one is made of stainless-steel and consists of several windings in order to keep the heat conductance as low as possible; then, a bobbin-shaped copper tube is secured to the 45-K plate; the third segment, identical to the first duct, minimizes thermal exchanges between the two PT stages; finally, a second spool-shaped copper pipe is fixed to the 4.2-K plate, intended to cool the helium gas down to a few K before entering the buffer cell. A second PID controller is also implemented for fine tuning of the buffer cell temperature. To keep the pressure within the radiation shields below $10^{-7}$ mbar, the internal surface of the inner shield is covered with a layer of activated charcoal that, at cryogenic temperatures, acts as a pump (with a speed of a few thousands dm$^3$/s) for helium and non-guided molecules; the gas adsorbed by the charcoal is released during warm up of the cryogenic system and then pumped out of the vessel by a turbomolecular pump. As shown in Fig. \ref{setup}, both the vacuum chamber, the shields and the buffer cell have optical accesses for spectroscopic interrogation. The probe radiation source is an external-cavity (Littman-Metcalf configuration) diode laser emitting several milliwatts of power between 1520 and 1570 nm with a linewidth below 1 MHz (New Focus, TLB-6300 Velocity). The laser output beam is split into four parts: one portion is sent to a room-temperature cell containing acetylene in order to identify the various transitions; a second beam is coupled to a confocal, Fabry-Perot (FP) interferometer for frequency calibration purposes; a third fraction is delivered to a wavelength meter with an accuracy of 0.2 ppm (Burleigh WA-1500); the last part passes through the buffer gas cell and is eventually collected by an InGaAs photo-detector (PD). The molecular absorption profile, $\delta(\nu)\equiv[I_0 -I(\nu)]/I_0$, is recovered by scanning the laser frequency $\nu$ through the application of a linear-ramp voltage to the piezoelectric transducer attached to the external-cavity tuning element.  

\section{Measurements and analysis}
\subsection{Translational temperatures}
\begin{figure}[h!]
\includegraphics[width=\textwidth]{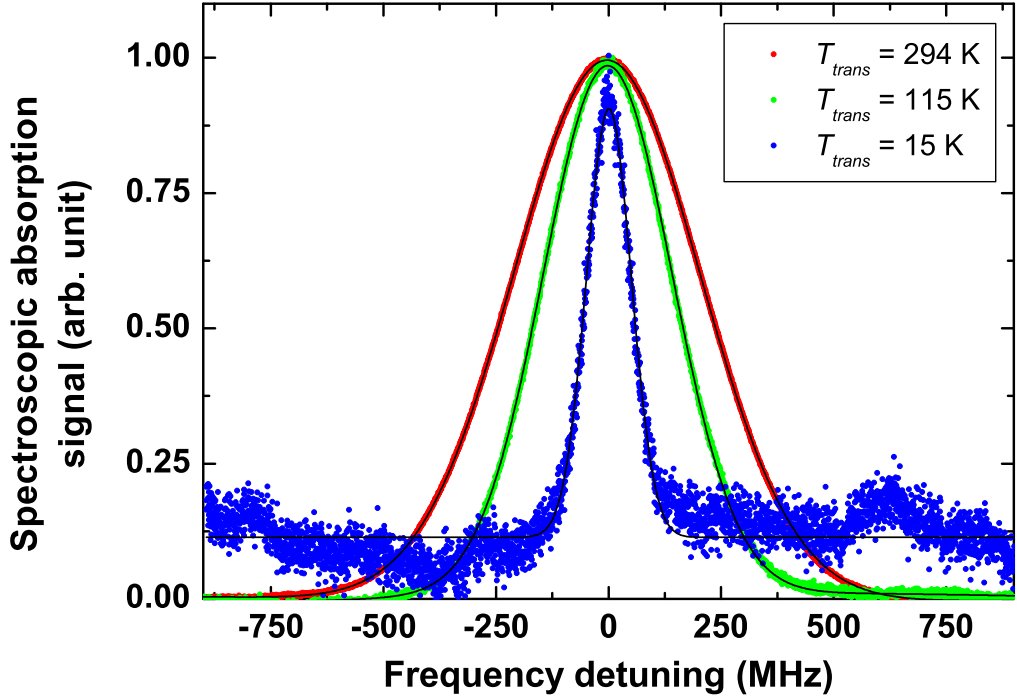}
\caption{Spectroscopic absorption signals (normalized to unit) obtained for transition $a$ in correspondence with the following triplets: $T_{cell}=294$, $f_{{\rm He}}=0$ SCCM, $f_{{\rm C}_2{\rm H}_2}=1$ SCCM; $T_{cell}=115$, $f_{{\rm He}}=20$ SCCM, $f_{{\rm C}_2{\rm H}_2}=5$ SCCM; $T_{cell}=10$, $f_{{\rm He}}=2$ SCCM, $f_{{\rm C}_2{\rm H}_2}=2$ SCCM. The extracted translational temperatures are: $294\pm 2$, $115\pm 5$, $15\pm 3$ K, respectively.}\label{larghezza}
\end{figure}

In order to monitor the collisional cooling process, the absorption spectrum of the R(5) ro-vibrational transition in the $(\upsilon_1 +\upsilon_3)$ band (henceforth referred to as transition $a$) was acquired under different experimental conditions, by varying the buffer-cell temperature, $T_{cell}$, and the two gas flows, $f_{{\rm He}}$ and $f_{{\rm C}_2{\rm H}_2}$. Since the translational temperature, $T_{trans}$, in a gas is related to the mean square velocity of its molecules, each observed absorption profile was fitted by a Gaussian distribution
\begin{equation}\label{gaussiana}
G(\nu)=G_0 \exp\left[-\frac{4\,\ln2\,(\nu-\nu_0)^2}{\sigma_{D}^2}\right]
\end{equation}
where the amplitude $G_0$, the transition center frequency $\nu_0$, and the Doppler width $\sigma_{D}=(\nu_0/c)\sqrt{8\ln2\,m^{-1}k_B T_{trans}}$ represent the fitting parameters (here, $m$ is the molecular mass, $c$ the light speed, and $k_B$ the Boltzmann constant). Thus, the translational temperature of the acetylene sample was retrieved by the extracted $\sigma_{D}$ value. As an example, three absorption spectra are shown in Fig. \ref{larghezza}, corresponding to the following $T_{cell}$ values: 294, 115 and 10 K; for $T_{cell}=294$ K, only 1 SCCM of acetylene was let into the cell and no helium; for $T_{cell}=115$ K, $f_{{\rm He}}=20$ SCCM and $f_{{\rm C}_2{\rm H}_2}=5$ SCCM were used; for $T_{cell}=10$ K, $f_{{\rm He}}=f_{{\rm C}_2{\rm H}_2}=2$ SCCM was found to be the optimal choice to reach the translationally coldest sample with our setup: $T_{trans}=15\pm 3$ K. Supported by a temperature reading of 15 K recorded on the He pipe just before the entrance into the buffer cell, the discrepancy at the lowest temperature was attributed to a non-perfect thermal exchange between the copper pipe and the two PT plates; to bridge this gap, an improved setup for better cooling of the He line is already under construction. It should be noted that equal flows of the two gases do not correspond to equal densities in the buffer cell. In fact, many of the acetylene molecules freeze upon impact on the walls (as well as on the optical windows), hence generating a layer of solid acetylene whose thickness increases with time. This is not the case for the helium. Nonetheless, after a short transient (less than 10 ms in the worst case), stationary gas densities, $n_{{\rm He}}$ and $n_{{\rm C}_2{\rm H}_2}$, namely gas pressures, will be established inside the buffer cell, leading to steady-state spectroscopic absorption profiles; these will eventually disappear as soon as the optical windows fog up. It is also worth remarking here that, in the work presented here, the stationary cell gas pressures were always lower than 0.2 mbar, giving rise to negligible pressure broadening and shift effects \cite{Bond2008}.

\subsection{Rotational temperatures} 
\begin{figure}[h!]
\centering
\includegraphics[width=\textwidth]{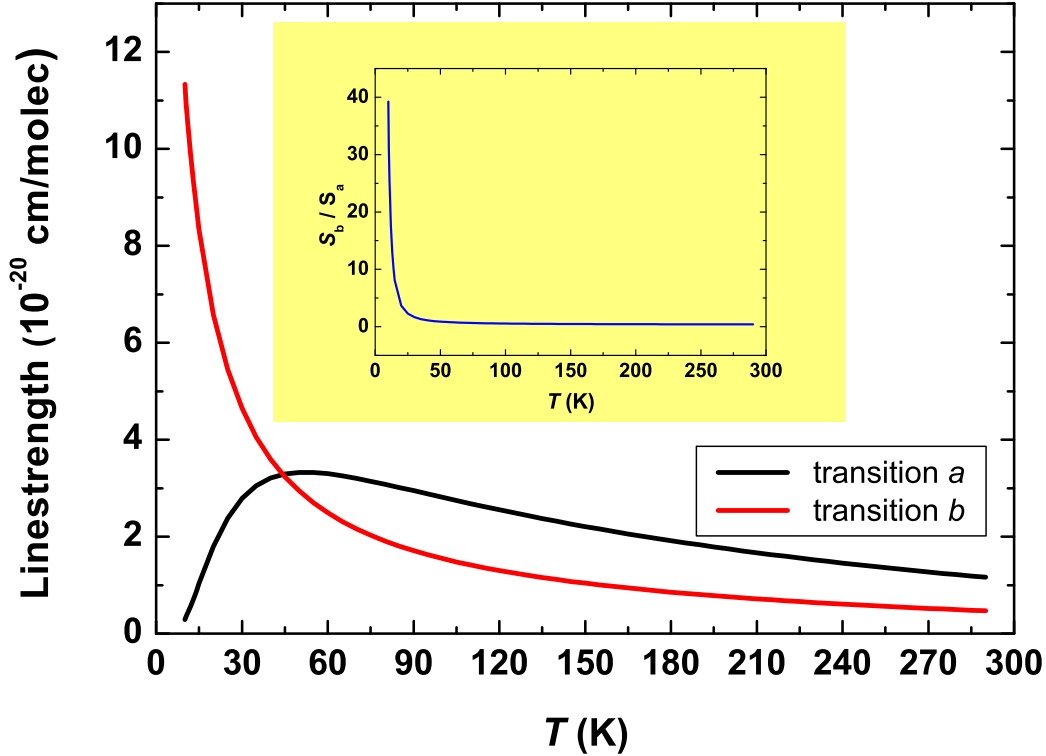}
\caption{Linestrengths of $a$ and $b$ transitions, calculated using Eq. \ref{line strength}, are plotted as functions of rotational temperature. The ratio between the two curves is plotted in the inset.}\label{linestrength}
\end{figure}

\begin{figure}[h!]
\centering
\includegraphics[width=\textwidth]{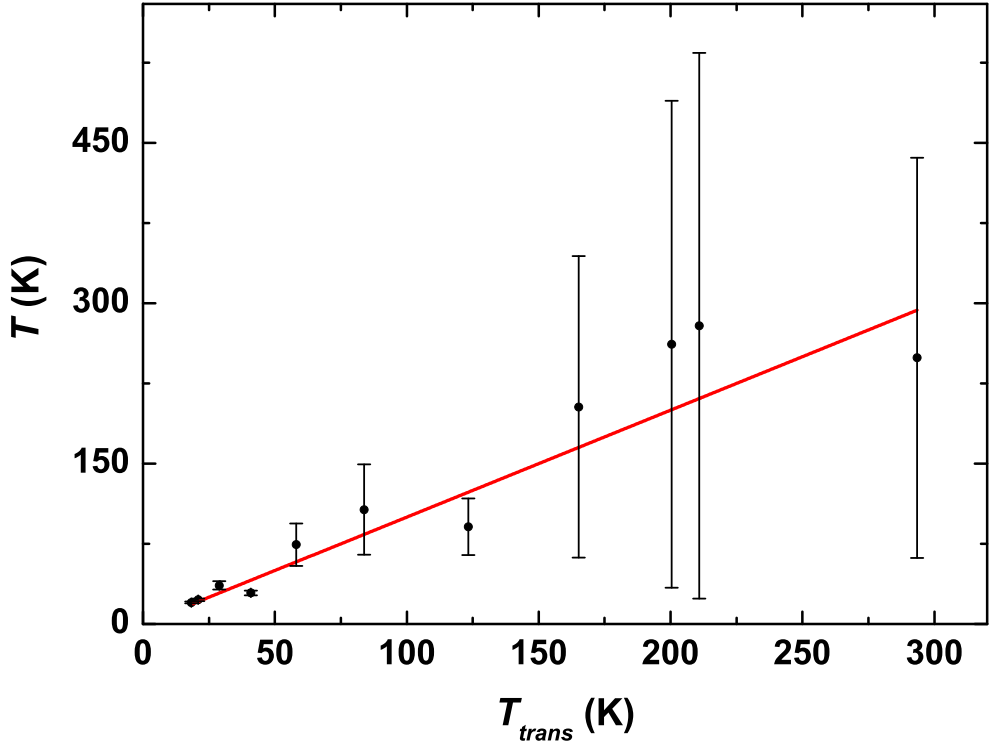}
\caption{Experimental rotational temperatures measured at different $T_{trans}$ values according to the procedure described in the text. The line $T=T_{trans}$ is also plotted for reference. It should be noted that each data point corresponds to a different choice of the two gas flows, essentially intended to maximize the signal-to-noise ratio of every absorption spectrum while reaching the lowest possible rotational temperature.}\label{trot}
\end{figure}

The linestrength of a given ro-vibrational transition depends on the rotational temperature, $T_{rot}$, hereafter simply called $T$ to simplify the notation, through the relationship \cite{Rothman1998}
\begin{equation}\label{line strength}
S(T)=S(T_{ref})\frac{Q(T_{ref})}{Q(T)}\frac{\exp\left(\frac{-c_2E_{f}}{T}\right)}{\exp\left(\frac{-c_2E_{f}}{T_{ref}}\right)}\frac{1-\exp\left(\frac{-c_2 \nu_{0}}{T}\right)}{1-\exp\left(\frac{-c_2 \nu_{0}}{T_{ref}}\right)}\ {\rm ,}
\end{equation}
where $T_{ref}$ is a reference rotational temperature at which the linestrength is known, $Q(T)$ the rotational partition function (varying between 3 at 10 K and 100 at 294 K in the case of acetylene \cite{Amyay2011}), $E_{f}$ the transition's lower-level energy (expressed in wavenumbers), and $c_2=hc/k_B$ ($h$ is the Plank constant). Eq. \ref{line strength} was exploited to perform accurate measurements of rotational temperatures according to the following procedure. First, besides transition $a$ (at $\nu_{0a}=6570.042687$ cm$^{-1}$), the $(\upsilon_1 +\upsilon_3)$ R(1) component, called transition $b$ (at $\nu_{0b}=6561.094106$ cm$^{-1}$), was also selected so that the ratio between the two respective linestrengths, 
\begin{equation}\label{line strength ratio}
R_{ba}(T)\equiv \frac{S_b(T)}{S_a(T)}=\frac{\frac{\exp\left(\frac{-c_2E_{fb}}{T}\right)}{\exp\left(\frac{-c_2E_{fb}}{T_{ref}}\right)}\frac{1-\exp\left(\frac{-c_2 \nu_{0b}}{T}\right)}{1-\exp\left(\frac{-c_2 \nu_{0b}}{T_{ref}}\right)}}{\frac{\exp\left(\frac{-c_2E_{fa}}{T}\right)}{\exp\left(\frac{-c_2E_{fa}}{T_{ref}}\right)}\frac{1-\exp\left(\frac{-c_2 \nu_{0a}}{T}\right)}{1-\exp\left(\frac{-c_2 \nu_{0a}}{T_{ref}}\right)}}\ {\rm ,}
\end{equation}
exhibits a steep slope below a few tens of Kelvin (see Fig. \ref{linestrength}), thus reducing errors in the determination of low rotational temperatures. Second, for different $T_{trans}$ values, the experimental value of $R_{ba}(T)\equiv S_b (T) /S_a (T)=\int\delta_b(\nu)\,d\nu /\int\delta_a(\nu)\,d\nu$ was determined. This value, along with the $E_f$'s and $\nu_0$'s parameters provided by the Hitran database \cite{Hitran}, was replaced in Eq. \ref{line strength ratio} which was finally solved for $T$ (see Fig. \ref{trot}). In conclusion, the minimum observed rotational temperature was $T=(20\pm 1)$ K for a measured translational temperature of $T_{trans}=(15\pm 3)$ K; such a difference is compatible with the fact that cooling is more efficient for the translational degrees of freedom than for rotational ones \cite{MaddaloniLibro}, albeit the two measured temperature values are consistent within 2 standard deviations.  

\begin{figure}[h!]
\centering
\includegraphics[width=\textwidth]{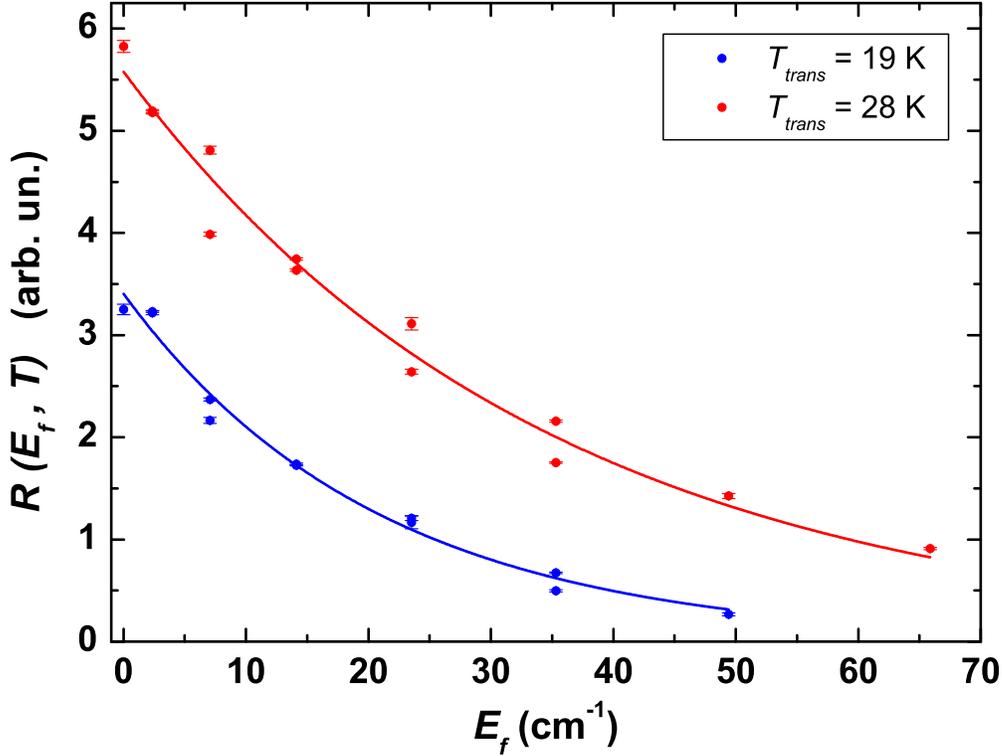}
\caption{Ascertainment of the canonical-invariance hypothesis. Eq. \ref{boltz} is fitted to the $R(E_f,T)$ data points measured as a function of the transition's lower-level energy for two different translational temperatures ($19\pm 2$ and $28 \pm 2$ K). The extraced rotational temperatures are $27\pm 2$ and $42 \pm 3$ K, respectively, with a fit correlation coefficient of $\chi=0.98$. Again, as in Fig. \ref{trot}, each data point is associated with a different pair of gas flows.}\label{moltelambda}
\end{figure}

In general, unlike what happens to translational states, even if the initial distribution over the rotational states is Boltzmannian, it will relax without preserving the canonical invariance, and it will not be possible to define a rotational temperature \cite{SannaLibro}. The necessary conditions so that the canonical invariance is maintained in a subsystem-reservoir relaxation process have been both mathematically and physically established \cite{Andersen1964}. To address this issue in our case, a normalized linestrength was measured for several ro-vibrational lines at a given translational temperature. The acquired behavior was then compared with the theoretical line dictated by the Boltzmann law. In practice, as shown in Fig. \ref{moltelambda}, the ratio $R(E_f,T)\equiv S(E_f, T)/S(E_f, T_{ref})=\int\delta(E_f,T,\nu)\,d\nu/\int\delta(E_f,T_{ref},\nu)\,d\nu$ was determined against $E_f$ (i.e., for each of the transitions listed in Table \ref{tabella}) in correspondence with two different $T_{trans}$ values, 19 and 28 K. It is worth pointing out that the normalization of $S(E_f, T)$ to $S(E_f, T_{ref})$ was necessary in order to get rid of the unknown dependence of $S(T_{ref})$ on $E_f$. The obtained data points were then fitted with the function
\begin{equation}\label{boltz}
R(E_f,T)=H \exp\left[c_2 E_f\left(-\frac{1}{T}+\frac{1}{T_{ref}}\right)\right]\ {\rm ,}
\end{equation}
with $H$ a proportionality constant, $T$ being the fitting parameter. The above equation is nothing but Eq. \ref{line strength} with $\left[1-\exp\left(\frac{-c_2 \nu_{0}}{T}\right)\right]\left[ 1-\exp\left(\frac{-c_2 \nu_{0}}{T_{ref}}\right)\right]^{-1}\simeq 1$. The extracted rotational temperatures were $T=(27\pm2)$ K and $T=(42\pm3)$ K for measured translational temperatures $T_{trans}=(19\pm2)$ K and $T_{trans}=(28\pm2)$ K, respectively. In both cases, the fit correlation coefficient was $\chi=0.98$, consistent with the hypothesis of canonical invariance.

\begin{table}[h]
\caption{\label{tabella}Frequencies and lower-level energies of the ro-vibrational lines used throughout this work, as provided by the Hitran Database \cite{Hitran}.}
\begin{indented}
\item[]\begin{tabular}{@{}ccc}
\br
$E_f$ (cm$^{-1})$ &$\nu_0$ (cm$^{-1}$) & \ $J$ and branch \\
\mr
35.2979  & 6570.042687  & R(5) $\equiv a$  \\
23.5323  & 6567.844393  & R(4)  \\
14.1195  & 6565.620174  & R(3)  \\
7.0598   & 6563.370066  & R(2)  \\
2.3533   & 6561.094106  & R(1) $\equiv b$   \\
0        & 6558.792333  & R(0)  \\
2.3533   & 6554.111497  & P(1)  \\
7.0598   & 6551.732512  & P(2)  \\
14.1195  & 6549.327869  & P(3)  \\
23.5323  & 6546.897607  & P(4)  \\
35.2979  & 6544.441767  & P(5)  \\
49.4163  & 6541.960389  & P(6)  \\
65.8871  & 6539.453516  & P(7)  \\
\br
\end{tabular}
\end{indented}
\end{table}

\subsection{Elastic cross section}
Finally, by comparing the measured diffusion time of $^{12}$C$_2$H$_2$ in the BGC cell (vs the $^{4}$He flux) with that predicted by a Monte Carlo simulation, we provided an estimate for the elastic cross section relevant to the translational cooling mechanism \cite{Lu2009, Skoff2011}. Let us start by looking a little more closely at the physics of the problem. After reaching thermal equilibrium with the He bath (under our typical experimental conditions, this happens on a path shorter than 100 $\mu$m, corresponding to about 50 collisions), a generic acetylene molecule experiences a random walk, scattered by helium atoms, until it freezes on the cell's walls or escapes through the exit hole (to form the molecular beam); in both cases, it stops contributing to the laser absorption. The larger the helium density, the higher the number of scattering events and the longer the acetylene average diffusion time, $\tau_{diff}$. 

This latter quantity was experimentally determined for transition $a$ via the relationship
\begin{equation}\label{tdiff}
\tau_{diff}(T_{trans})=\frac{L_c^3\,n_{{\rm C}_2{\rm H}_2}(T_{trans})}{f_{{\rm C}_2{\rm H}_2}}=\frac{L_c^3}{f_{{\rm C}_2{\rm H}_2}}\frac{\sigma_D (T_{trans})\,\int\delta_a(\nu, T_{trans})d\nu}{S_a(T)\,L_c}\ {\rm ,}
\end{equation}
where the spectroscopic derivation of the acetylene density (based on the Lambert-Beer law) was also used \cite{MaddaloniLibro}. Then, the $S_a(T)$ value corresponding to the measured $\sigma_D (T_{trans})$ was calculated by means of Eq. \ref{line strength}, with $E_f$, $\nu_0$, and $S_a(T_{ref}=294\ {\rm K})=1.13\cdot 10^{-20}$ cm/molec taken from the Hitran Database, and $Q(T)$ provided by Amyay and coworkers \cite{Amyay2011}. It should be noted that, in the above procedure, $T=T_{trans}$ was inevitably assumed. To fix that, curve $a$ in Fig. \ref{linestrength} was used to estimate the extent to which the discrepancy between $T$ and $T_{trans}$ (as measured in Fig. \ref{trot}) affects the determination of $S_a (T)$; this is reflected in conservatively augmented error bars on the $\tau_{diff}$ data points. 

The helium density was derived though the formula \cite{Hutzler2012}
\begin{equation}\label{tdiff}
n_{{\rm He}}=\frac{4f_{{\rm He}}}{\pi r_h^2\left \langle v_{{\rm He}} \right\rangle}\ {\rm ,}
\end{equation}
with $\left\langle v_{{\rm He}}\right\rangle =\sqrt{8k_BT_{trans}\pi^{-1}m_{{\rm He}}^{-1}}$ being the mean thermal velocity of helium particles ($m_{{\rm He}}$ is the helium atom mass). According to the above procedure, two sets of $\tau_{diff}$ vs $n_{{\rm He}}$ were recorded, corresponding to translational temperatures of 100 and 25 K, respectively. 

\begin{figure}[h!]
\centering
\includegraphics[width=\textwidth]{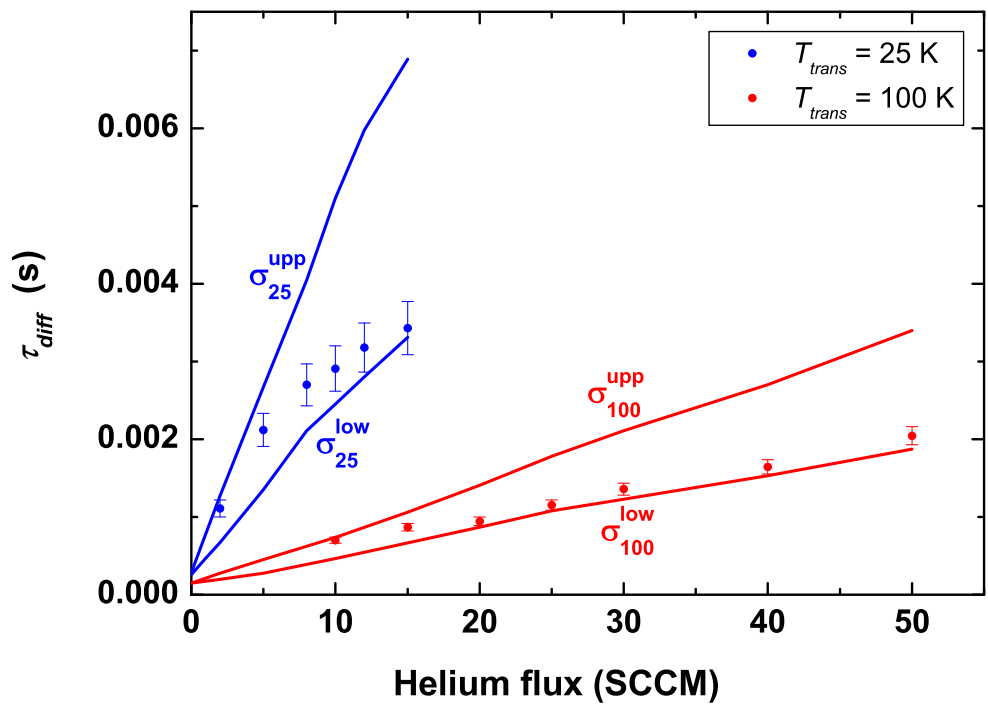}
\caption{Experimental acetylene diffusion time plotted against $f_{{\rm He}}$ at a constant acetylene flux: $f_{{\rm C}_2{\rm H}_2}=5$ SCCM for $T_{trans}=25$ K and $f_{{\rm C}_2{\rm H}_2}=$50 SCCM for $T_{trans}=100$ K. Theoretical simulations (continuous lines) are also shown which delimit the measured data from above and from below ($\sigma_{100}^{upp}=9.0\cdot 10^{-20}$ m$^{2}$, $\sigma_{100}^{low}=4.6\cdot 10^{-20}$ m$^{2}$; $\sigma_{25}^{upp}=4.9\cdot 10^{-20}$ m$^{2}$, $\sigma_{25}^{low}=3.1\cdot 10^{-20}$ m$^{2}$), thus enabling the estimate of the total elastic cross sections.}\label{simulazione}
\end{figure}

In a second stage, a theoretical simulation was carried out to reproduce the measured acetylene diffusion times. In particular, the $^4$He$-^{12}$C$_2$H$_2$ interaction was processed by a conventional Monte Carlo method, whereas the molecule free evolution was made to follow Newton's law. Firstly, for a given translational temperature, an acetylene molecule was injected into the buffer cell at time $t=0$ with its three velocity components extracted randomly according to the corresponding Maxwell-Boltzmann distribution. Then, the probability $\mathcal{P}$ that an interaction occurs in the elementary interval time $\delta t$ was calculated considering the $^4$He$-^{12}$C$_2$H$_2$ relative velocity ($v_{rel}$), $n_{{\rm He}}$, and a trial cross section ($\sigma_{tr}$): $\mathcal{P}=n_{{\rm He}}\,\sigma_{tr}\,v_{rel}\,\delta t$. After that, a random number $\mathcal{N}$ (between 0 and 1) was generated: if $\mathcal{N}<\mathcal{P}$, then the atom-molecule impact was allowed to take place and new random velocity components were consequently extracted; otherwise, the molecule evolved freely for a successive time interval $\delta t$. These steps were iterated for successive $\delta t$ intervals until the molecule reached one of the walls: the time $\tau$ spent in the cell before freezing was accordingly calculated. The whole procedure was then repeated for a thousand injected molecules, namely the minimum allowed number which doesn't affect the simulation result; averaging over all the computed $\tau$ values eventually yielded $\tau_{diff}$. Depending on the values of $f_{{\rm He}}$, $T_{trans}$ and $\sigma_{tr}$, a different time interval $\delta t$ was used in the simulation. Its value was kept between 1 and 10 ns, i.e. always small enough not to alter the simulation outcome. For each of the two translational temperatures, the above simulation was carried out as a function of $n_{{\rm He}}$, searching for the optimal pair of $\sigma_{tr}$ values which strictly delimits the experimental points from above and from below. The results are shown in Fig. \ref{simulazione}. The elastic cross sections were estimated to be $\sigma_{el}(T_{trans}=100\ {\rm K})=(4\pm1)\cdot 10^{-20}$ m$^{2}$ and $\sigma_{el}(T_{trans}=25\ {\rm K})=(7\pm2)\cdot 10^{-20}$ m$^{2}$.

\section{Conclusion}
Thanks to the implementation of a modern buffer-gas-cooling technique, the range of cryogenic temperatures for the spectroscopic study of acetylene in a helium environment was extended, with respect to previous literature, down to a few Kelvin. In order to accurately determine the achieved translational and rotational temperatures, several ro-vibrational transitions belonging to the ($\upsilon_1 +\upsilon_3$) band were used for Doppler thermometry and measurements of relative intensities. The attainment of a well-defined rotational temperature for the considered $^4$He$-^{12}$C$_2$H$_2$ system was also demonstrated. Finally, a deeper insight into the collisional cooling process was gained by measuring the acetylene diffusion time in the buffer cell against the helium density at two temperatures that spanned a large range (100 and 25 K); in this respect, an appropriate theoretical model was also developed, which allowed us to obtain an estimate for the respective elastic cross sections. These figures may be particularly useful in planetary science when modeling the process of translational energy relaxation of molecules in bath gases, which is crucial for understanding the energy balance of the upper atmosphere and its evolution \cite{Bovino2011, Nan1992}.  

While insignificant in the range of pressures explored in this work, pressure broadening and shifts are also of foremost importance at temperatures of astrophysical relevance and, as such, will be the subject of future investigations. Moreover, accurate analysis/modeling of spectral lineshapes represents a powerful tool for probing fundamental atom-molecule low-temperature interaction processes; obtaining molecular spectra with enhanced signal-to-noise ratios is vital for addressing this issue and, indeed, work is in progress for the implementation of a cavity ring-down spectroscopy technique. Finally, by virtue of the enormous versatility of our buffer-gas-cooling apparatus, the spectroscopic study reported here may be readily extended to other fundamental atmospherical and astrophysical molecular species such as, for instance, methane \cite{Onstott2006, Lellouch2009}, nitrous oxide \cite{Ziurys1994}, and carbon dioxide \cite{Oancea2012}.

\vspace{2pc}
The authors acknowledge technical support by G. Notariale. This work was funded by MIUR-FIRB project RBFR1006TZ and by INFN project SUPREMO.

%

\end{document}